\documentclass[aps,prl,twocolumn,groupedaddress]{revtex4}

\usepackage{graphicx}
\usepackage{dcolumn}
\usepackage{bm}
\usepackage{longtable}
\usepackage{xcolor,soul}
\sethlcolor{yellow}

\begin{document}


\title{The Phase Diagram for Percolating Free Surfaces in Disordered Assemblies of 
Faceted Grains} 


\author{D. J. Priour, Jr}
\affiliation{Department of Physics \& Astronomy, Youngstown State University, Youngstown, OH 44555, USA}
\altaffiliation{}


\date{\today}

\begin{abstract}
Percolation in systems made up of randomly placed impermeable grains is often examined 
in the context of system spanning clusters of connected solids forming above a 
relatively low critical grain density $\rho_{c1}$ or networks of interstitial void volumes
ceasing to exist above a signficantly higher threshold $\rho_{c2}$.  In this work, we interpret these 
percolation transitions as, respectively, the low and high density boundaries of percolating 
exposed surfaces which either ensheath clusters of impermeable particles or line tunnel-like voids.  
Moreover, we find in the thermodynamic limit exposed surfaces are either sheaths 
or tunnels with a second order phase transition from the former to the latter at a density threshold 
$\rho_{c*}$) intermediate between $\rho_{c1}$ and $\rho_{c2}$. 
We calculate critical inclusion densities with a new method which identifies exposed free surfaces in a 
geometrically exact manner with a computational cost scaling only linearly in the system volume.
We obtain $\rho_{c1}$, $\rho_{c2}$, and $\rho_{c*}$ for a variety of grain geometries, including each of the 
Platonic solids, truncated icosahedra, and structurally disordered inclusions formed from cubes subject to a random
sequence of slicing planes. In the case of the latter, we find a limiting value of $5\%$ for the critical
porosity at the void percolation threshold as the number of sustained slices per cube becomes large.
\end{abstract}

\maketitle
Though well characterized in discrete lattices~\cite{stauffer}, percolation phenomena in disordered continuum systems 
such as assemblies of barrier particles impermeable to fluid flow or charge transport often lack a well 
defined lattice; instead, 
transport characteristics (e.g involving charge or fluid)  are strongly influenced by the number density 
$\rho$ per unit volume of impermeable barrier particles or grains making up the system.  
Theoretical studies of percolation transitions in such systems often examine either 
system spanning clusters of overlapping particles (for which we use the label $\rho_{c1}$ for both the 
transition and its critical density) 
or networks of interstitial void volumes navigable 
on macroscopic scales (henceforth referred to as $\rho_{c2}$ both in terms of the critical concentration and 
the transition itself).     
While connected networks of overlapping grains form at relatively low $\rho_{c1}$ values with the bulk of the system 
consisting of empty space, void networks persist to much higher $\rho_{c2}$ values where the interstitial 
volumes represent only a small portion of the system volume as a whole.

In general, a 
geometrically exact description of the void volumes themselves has proven elusive,
though discretization schemes has been applied in some cases~\cite{martys,maier,Yi1,Yi2,koza} 
and Voronoi networks~\cite{elam,marck,rintoul,klatt}, 
have been brought to bear in the 
case of systems made up of randomly placed spheres. Simulations involving virtual tracer particles infiltrating 
interstitial volumes have been used to calculate critical parameters such as 
$\rho_{c}$~\cite{bauer,beijeren,hofling,spanner,hofling2,kammerer,spanner2,djpriour,djpriour2,ballow}, but have not
elucidated the irregularly shaped voids themselves.  

In this work, we develop a technique to 
determine surfaces lining both grain clusters and tunnel-like interstitial volumes in a geometrically 
exact manner with a computational cost scaling only linearly in the system volume. 
We calculate $\rho_{c1}$ and $\rho_{c2}$ for a variety of grain geometries.  In the case of the latter 
(i.e. void percolation), accuracy gains in the critical densities we report on here reveal 
differences in $\rho_{c2}$ for aligned versus randomly oriented grains, which apart from the case of cube 
shaped grains had not previously been resolved.  Moreover, as an example of structurally disordered grains, 
we consider the case of irregular fragments formed by subjecting 
cubes to a succession of randomly oriented fracturing planes; we find  
the  critical porosity fraction to be $5 \%$ after many accumulated slices,

We consider the formation of worm-like clusters of overlapping grains just above $\rho_{c1}$ and the 
emergence of system spanning void networks for $\rho$ decreasing below $\rho_{c2}$ 
to be, respectively, the low and high density limit of a 
phase of percolating free surfaces.  
A recent work discusses this regime in terms bicontinuous components of a two phase 
material.  The emphasis in this work, apart seeking to improve the accuracy of $\rho_{c2}$ (void 
percolation critical concentrations) is 
to further characterize the percolating free surfaces for $\rho_{c1} < \rho < \rho_{c2}$.

Qualitatively speaking, one may be inclined to think of exposed surfaces 
as mainly sheaths for $\rho$ in the vicinity of $\rho_{c1}$ and as tunnels lining interstitial void 
at much high inclusion densities near $\rho_{c2}$. 
To locate the transition from sheath-like to 
tunnel-like surfaces, we define an order parameter to test whether 
exposed surfaces are sheaths or tunnels; for this purpose 
we decrease the linear size of each impermeable grain by a
very small fraction, 
recalculate the exposed surface, and then determine if it expands (indicative of a tunnel as the void 
volumes grow larger) or contracts (as would be anticipated for a sheath bounding a shrinking cluster of 
overlapping grains). In particular, we calculate the resulting difference $\Delta_{\mathrm{rms}}$ in the RMS 
displacement of vertices from their mean positions to determine if a surface is a sheath or a tunnel.
Using this quantitative measure, we identify a third phase boundary for a critical concentration (labeled 
in this work as $\rho_{c*}$) intermediate between $\rho_{c1}$ and $\rho_{c2}$ where percolating free surfaces are sheaths for 
$\rho < \rho_{c*}$ and tunnels for $\rho > \rho_{c*}$.  While this transition exhibits all the hallmarks of a second order 
phase transition, we find its critical behavior to be distinct from that of the standard 3D 
percolation universality class.

To minimize finite size effects and to circumvent artifacts in which spurious large boundary surfaces form above 
$\rho_{c1}$ in the case of a free boundary, we use periodic boundary conditions.
For the sake of computational efficiency, we 
partition the simulation volume into small cube shaped voxels (similar to verlet cells~\cite{bruin}); each 
of the small cells, with edges of unit length, is populated with impermeable grains by sampling Poissonian statistics.
The constituent grains are circumscribed by  
spheres of unit radius, though they need not be tangent to the sphere.
Each face of the polyhedral grains is defined by a plane, and
with vertices being formed by the intersection of three planes, we concentrate on these component planes, bounded by the 
polygonal faces of the parent polyhedra.  
As is described in greater detail elsewhere~\cite{djpriour2}, the planes of randomly 
oriented solids are described in terms of stochastically generated orthogonal axes.

We find the vertices lining exposed surfaces by proceeding in three stages.  First, based on the geometry of 
the host polyhedron, one identifies the edges of each face even as 
 planes engulfed entirely by individual neighboring polyhedra are 
eliminated from further consideration.  Next, by sweeping over nearby voxels, one finds each plane's  neighbors, 
a condition met if the edges of either polygonal face pass through the interior of the other.  Finally, one considers all 
possible sets of three planes which could form a vertex candidate for an exposed surface.  For each plane in the system, one 
sweeps through secondary and tertiary planes sampled from the neighbors established in the previous step with combinatorial 
counting preventing redundant consideration of sets of three planes.  
Only candidates lying on the parent 
polyhedra of the three planes forming the vertex while not being interior to any of the host polyhedra of 
neighboring planes are vertices on an exposed surface.

Next, we identify all vertices which belong to a free surface and only that free surface.
From each point emanate three edges which terminate in neighboring vertices, which are needed to work 
out the (often non-convex) polygonal faces of surface.  With each edge being defined by the intersection 
of two of the three planes forming the vertex, the neighboring point on the edge is the closest vertex belonging 
to both edge planes such that the edge extending to the point does not cross an empty chasm or pass 
through an impermeable grain.

Finally, one enumerates all connected polygons inscribed in the faces of host polyhedra by choosing a point on the plane, 
moving among nearest neighbors by traversing only edges confined to the plane until the ring closes on itself.  This 
process, similar in spirit to the Hoshen-Kopelman algorithm~\cite{hoshen}, is iterated until no unassigned points remain.
One may be tempted to identify polygons obtained in this way with the faces of an exposed surface, but this is only  valid 
if the face is topologically simple, a condition not met (e.g.) if the corner of a grain penetrates the face of 
another inclusion without contacting any edges; 
we therefore need to keep track of the structure of the faces of the
free surface.  
For a given polygon, we do this by first finding the nearest circumscribing polygon, if any exist,  
by choosing a point at random on a segment of the polygon under consideration and constructing a planar 
ray to the boundary of the face of the host polyhedron.  
One distinguishes among circumscribing and non-circumscribing rings by appealing to the number of ray crossings, 
odd for the former and even for the latter.
The internal polygon and the nearest
external candidate are part of the same face if in traversing the gap between the two rings the
ray does not pass through an impermeable grain or traverse a chasm of empty space.  
Continuity of the surface as a connected network of vertices is then maintained by choosing at random  
points in both the inner and outer polygons and making them each other's neighbors.

We find the additional step of associating faces makes a significant difference in $\rho_{c1}$ calculations (with the exception 
of aligned mono-dispersed cube-shaped grains where faces of exposed surfaces are invariably topologically simple), while
$\rho_{c2}$ results are not affected to even the slightest degree in any instance 
we have considered.

\begin{figure}
\includegraphics[width=.4\textwidth]{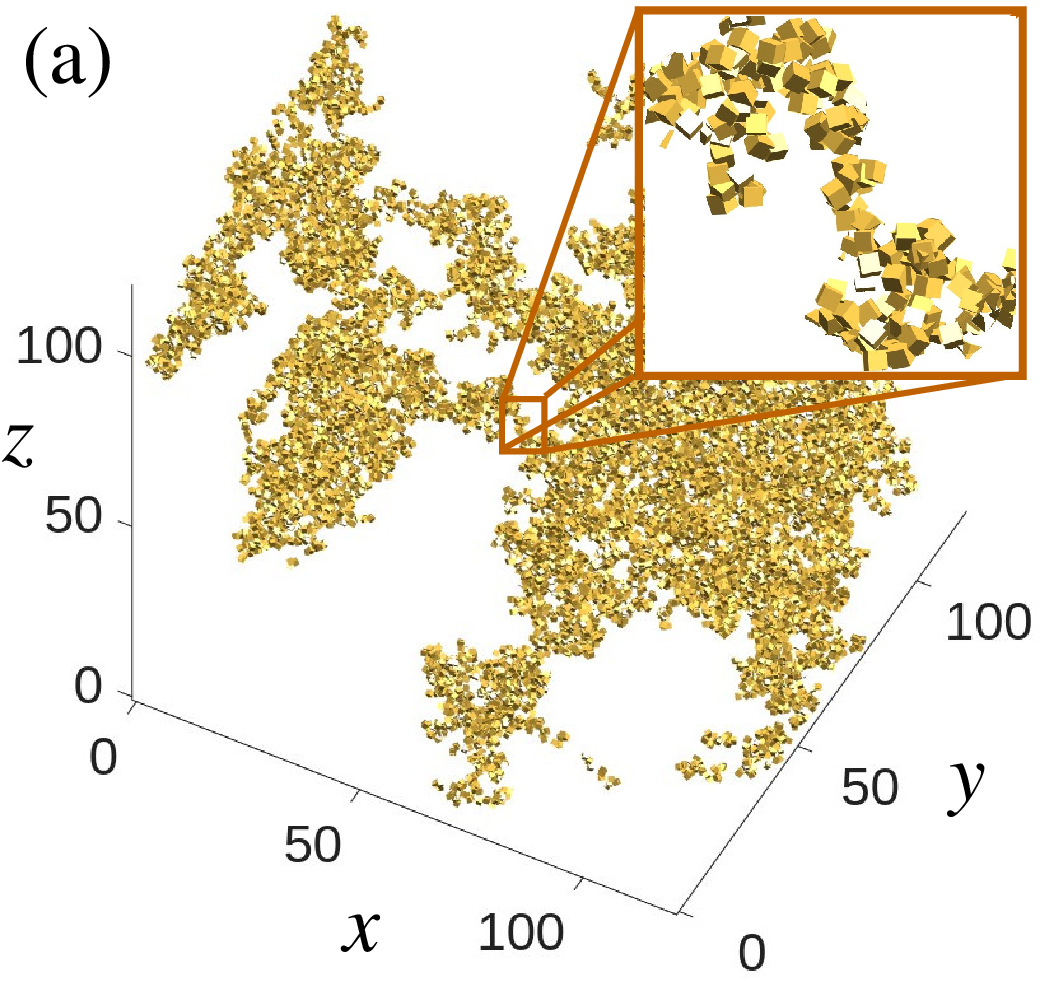}
\includegraphics[width=.4\textwidth]{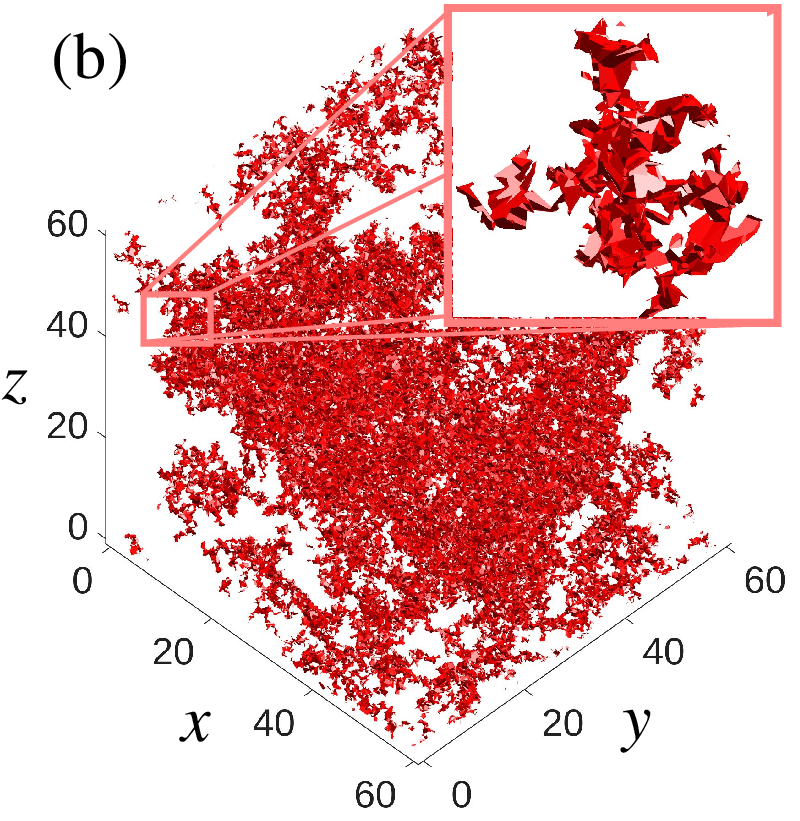}
\caption{\label{fig:Fig1} (Color online) Percolating free surfaces in the
case of randomly oriented cube shaped grains shown near $\rho_{c1}$ in panel (a) and near
$\rho_{c2}$ in panel (b).}
\end{figure}

Based on the effective size  $L_{\mathrm{eff}} \equiv L \rho^{1/3}$ (with $L$ being the linear dimension of the simulation 
volume in voxels), the foregoing technique can analyze identify all free surfaces in a  $L_{\mathrm{eff}} \sim 40$ 
system in typically less than a minute on a single computational thread within a contemporary CPU.  
All results reported
in this work involve averaging over at least $2 \times 10^{4}$ realizations of disorder.
In the case of the Platonic solids, where the accuracy standard in this work is one part in $10^{3}$ or 
better, we consider system sizes up to at least $L_{\mathrm{eff}} = 40$.
Free surfaces near percolation thresholds $\rho_{c1}$ and $\rho_{c2}$ are shown in Fig.~\ref{fig:Fig1} in panel (a) and
panel (b) respectively for the case of randomly oriented cubes.

Our grain-grain cluster percolation and void percolation
 results are based on calculations of $\langle f \rangle$, the 
disorder averaged percolation fraction.
To determine if a free surface percolates, in a given Cartesian direction, we count as 
system spanning a surface which wraps around (extending the length of the simulation volume and 
connecting with itself) or extends the length of the 
system without necessarily joining with itself.  
We exploit the inherent subjectivity of the latter to reduce corrections to leading order finite size scaling by
introducing a tunable parameter $\chi$ such that the condition for percolation in ,e.g., the $x$ direction is 
$(x_{\mathrm{max}} - x_{\mathrm{min}}) + \chi \geq L$.   
This percolation condition is imposed simultaneously in the $x$, $y$, and $z$ directions.
We optimize $\chi$ by insisting that $\rho_{c}$ results for intersections of $\langle f \rangle$ for 
distinct pairs of system sizes are the same. 

\begin{figure}
\includegraphics[width=.45\textwidth]{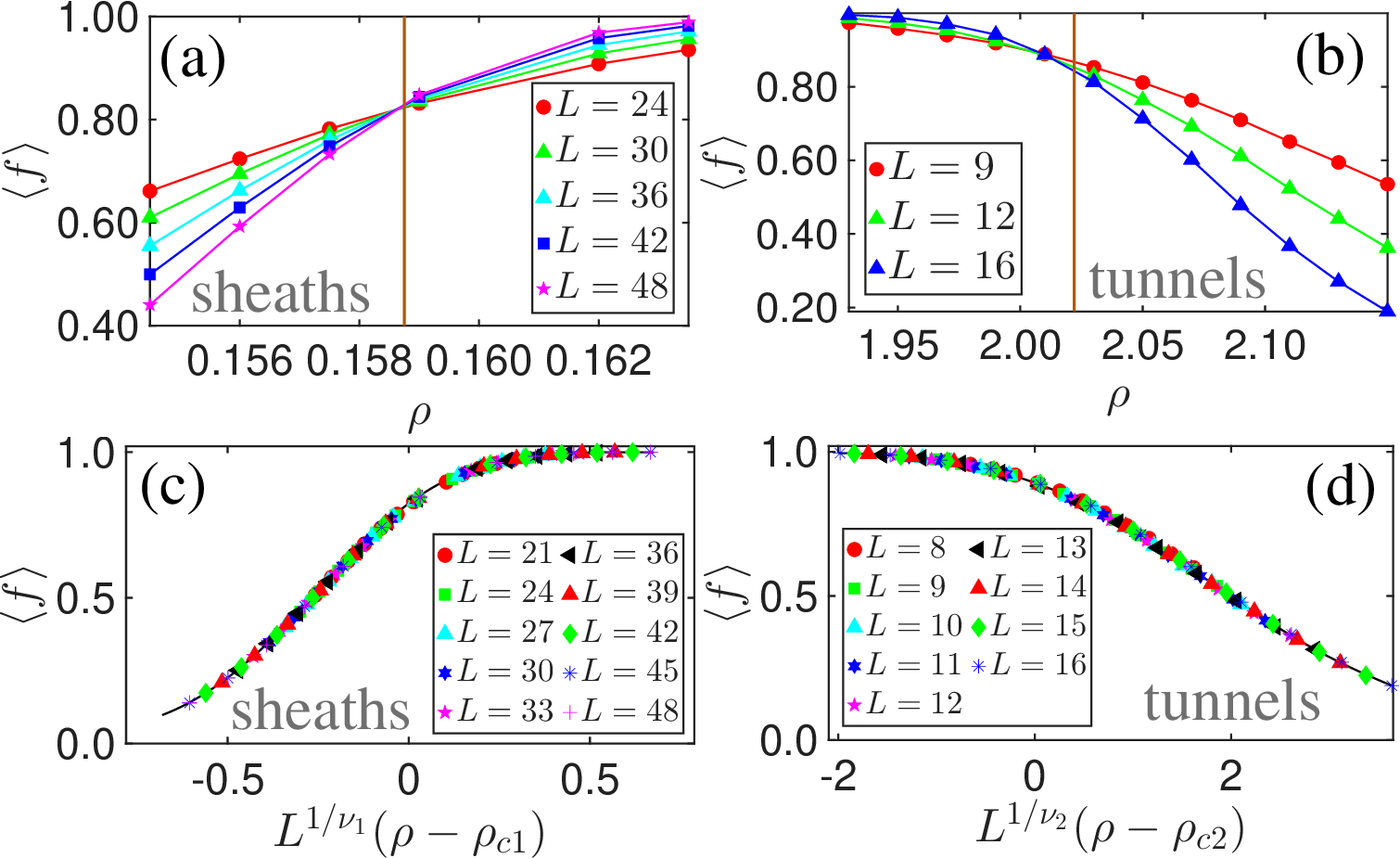}
\caption{\label{fig:Fig2} (Color online) $\langle f \rangle$ for $\rho_{c1}$ and $\rho_{c2}$ are shown in 
panel (a) and (b), respectively for the case of randomly oriented cubes; panels (c) and (d) show the corresponding data collapses.}
\end{figure}

We calculate critical concentrations by comparing $\left \{ L, \frac{3}{2} L \right \}$ pairs for 
multiple $L$ values and fitting to $\rho_{c} = \rho_{c0} + A L^{-\delta}$ with $A$ minimized by optimal choice 
of $\chi$.  In addition, we also perform quantitative data collapses, as described in a previous work.~\cite{djpriour};  
we find through least squares fitting the
polynomial $g(x) = \sum_{j = 1}A_{j} x^{j}$ (where $x = L^{1/\nu} [\rho - \rho_{c}]$) to Monte Carlo data
with $\nu$ and $\rho_{c}$ tuned to reduce the chi square deviation measure. Up to Monte Carlo statistical errors, 
$\left \{ L, \frac{3}{2} L \right \}$ $\langle f \rangle$ intersections and quantitative deta collapse yield 
identical critical density results.

Fig.~\ref{fig:Fig2} shows $\langle f \rangle$ for $\rho_{c1}$ and $\rho_{c2}$ in panel (a) and panel (b) 
respectively for randomly oriented cubes.  The vertical lines indicate critical concentrations obtained in this work for 
significantly greater system sizes.  Panel (c) and panel (d) show the corresponding quantitative data 
collapses, with the solid line being the previously mentioned analytical scaling curve.

\begin{figure}
\includegraphics[width=.45\textwidth]{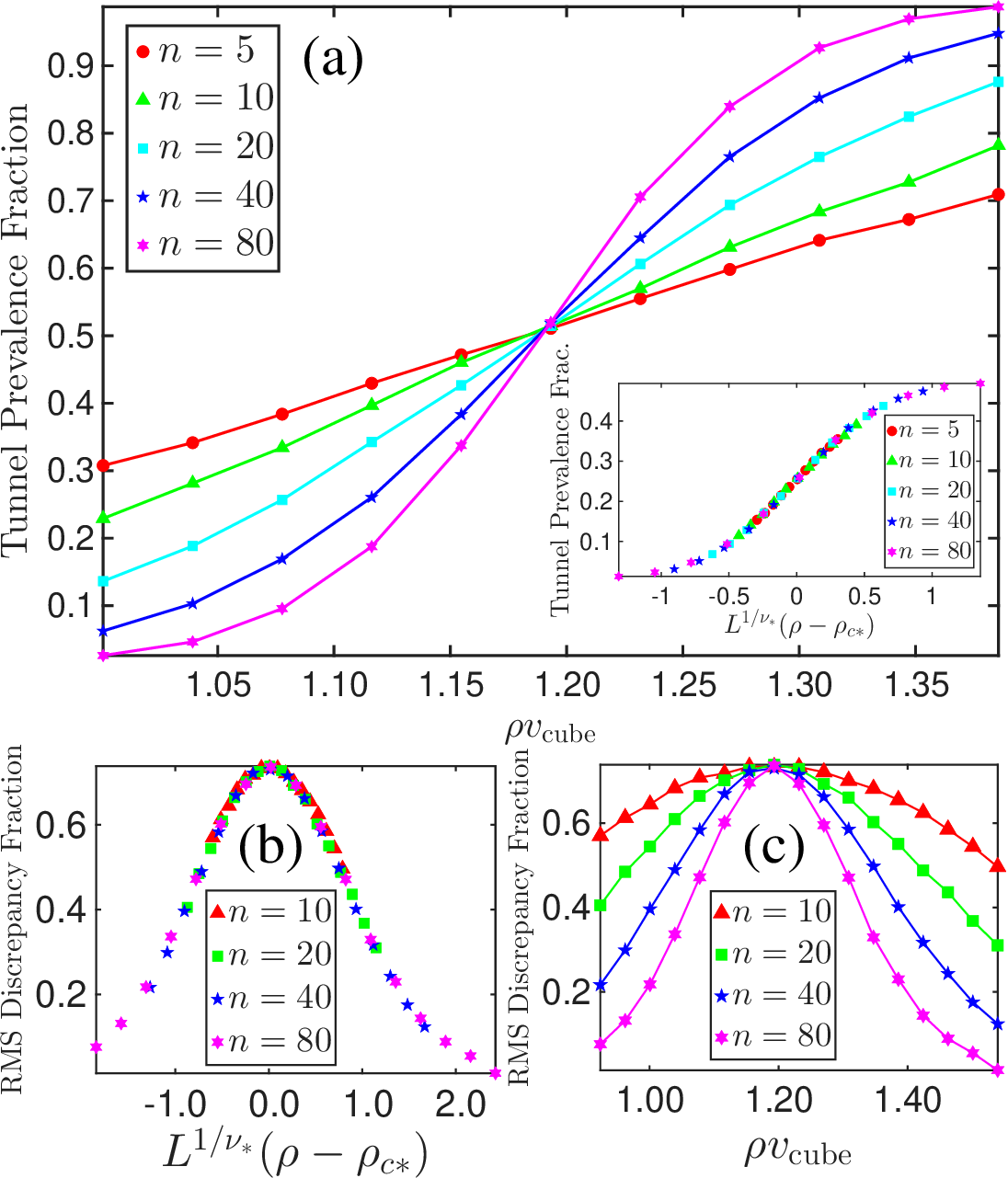}
\caption{\label{fig:Fig3} (Color online) Sheath/tunnel transition results are shown for the 
case of randomly oriented cubes.  The main plot in panel (a) displays the tunnel prevalence parameter for 
various system sizes, while the inset shows the same results in a data collapse graph.  The RMS discrepancy 
parameter is shown in panel (b), and is plotted in the form of a data collapse in panel (c).}
\end{figure}

As a quantitative measure of whether a free surface is a tunnel or a sheath,  
we calculate the difference in root mean square
(RMS) displacements $\Delta_{\mathrm{rms}}$ 
of vertices from their mean position prior to and after a small reduction (i.e. one  part in $10^{7}$) of 
the linear dimensions of the faceted grains.  
Anticipating an inverse relationship in the change of the size of the void volumes and that of the 
impermeable grains, one would expect an expansion of free surfaces lining tunnels to 
accompany a reduction in grain sizes;
we therefore take the frequency of 
positive $\Delta_{\mathrm{rms}}$ (weighted by the number of vertices in the free surface) to be a measure of the 
prevalence of tunnels.  The disorder averaged 
tunnel prevalence parameter is shown in panel (a) of Fig.~\ref{fig:Fig3} 
for randomly oriented cubes for a variety of system sizes with  
a well defined common crossing at $\rho_{c*}$.  The inset of panel (a) shows a collapse of the points 
main plot onto a common scaling curve when plotted with respect to $L^{1/\nu_{*}} (\rho - \rho_{c*})$ for
$\nu_{*} = 1.85$ and $\eta_{c*} = 1.19$.

Alternatively, one may examine the frequency of discrepancies of $\Delta_{\mathrm{rms}}$ components
for the $x$, $y$, and $z$ directions. The disorder averaged discrepancy parameter, shown
for various system sizes in panel (b) of Fig.~\ref{fig:Fig3},  peaks at $\rho_{c*}$ and 
becomes narrower with increasing $L$.  The data collapse in panel (c) of Fig.~\ref{fig:Fig3} 
for the same $\rho_{c*}$ and $\nu_{*}$ as in the case of the inset of panel (a) indicates that the discrepancy 
parameter is governed by the same critical indices as the mean tunnel prevalence fraction. 

\begingroup
\squeezetable
\begin{table}[h]    
\centering
\begin{tabular}{| c | c | c | c | c | c | c | c |}
\hline
\hline
$Grain$ & $\eta_{c1}$ & $\nu_{1}$ & $\eta_{c*}$ & $\nu_{*}$ & $\eta_{c2}$ & $\nu_{c}$  \\
\hline
$\textrm{Tet}_{\mathrm{A}}$ & 0.13436(5) & 0.90(4) & 0.968(3) & 1.4(1) & 2.836(2) & 0.88(5) \\
$\textrm{Tet}_{\mathrm{R}}$ & 0.16643(5) & 0.87(5) & 0.862(3) & 1.8(1) & 2.830(2) & 0.96(9) \\
$\textrm{Cub}_{\mathrm{A}}$ & 0.3247(2) & 0.89(2) & 1.80(2) & 1.25(15) & 3.279(2) & 0.90(5) \\
$\textrm{Cub}_{\mathrm{R}}$ & 0.2445(2) & 0.93(3) & 1.190(5) & 1.85(15) & 3.113(2) & 0.87(5) \\
$\textrm{Oct}_{\mathrm{A}}$ & 0.3272(1) & 0.88(4) & 1.75(1) & 1.25(15) & 3.207(2) & 0.77(13) \\
$\textrm{Oct}_{\mathrm{R}}$ & 0.2517(3) & 0.88(3) & 1.433(7) & 1.80(20) & 3.252(3) & 0.99(18) \\
$\textrm{Dod}_{\mathrm{A}}$ & 0.3385(1) & 0.88(3) & 1.615(8) & 1.25(15) & 3.348(2) & 0.85(7) \\
$\textrm{Dod}_{\mathrm{R}}$ & 0.2987(1) & 0.90(2) & 1.448(8) & 1.65(15) & 3.339(2) & 0.83(8) \\
$\textrm{Ico}_{\mathrm{A}}$ & 0.3393(5) & 0.89(5) & 1.669(8) & 1.25(15) & 3.381(3) & 0.87(7) \\
$\textrm{Ico}_{\mathrm{R}}$ & 0.3054(3) & 0.85(5) & 1.621(8) & 1.80(15) & 3.414(3) & 0.91(5) \\
$\textrm{Tr Ico}_{\mathrm{A}}$ & 0.3414(2) & 0.88(4) & 1.664(7) & 1.20(15) & 3.459(2) & 0.91(5)  \\
$\textrm{Tr Ico}_{\mathrm{R}}$ & 0.3263(2) & 0.88(4) & 1.599(7) & 1.70(15) & 3.452(2) & 0.91(6) \\
\hline
\hline
\end{tabular}
\caption{\label{tab:Tab1} Critical grain concentrations and critical exponents for $\eta_{c1}$, 
$\eta_{c*}$, and $\eta_{c2}$.  Three letter abbreviations (e.g. ``Cub'' for cubes) indicate
the grain shapes, while subscripts ``A'' and ``R'' indicate aligned and randomly oriented cases 
respectively.} 
\end{table}
\endgroup
Table~\ref{tab:Tab1} displays critical concentrations and critical exponents for the $\eta_{c1}$,
$\eta_{c*}$, and the $\eta_{c2}$ transitions where $\eta_{c} = \rho_{c} v_{\mathrm{B}}$ with 
$v_{\mathrm{B}}$ being the inclusions volume; volume results for the Platonic solids and the 
semi-regular truncated icosahedron may be found in~\cite{polyvolume}. 
Apart from the case of randomly oriented tetrahedra, our $\eta_{c1}$ results are in good general agreement with 
results previously reported in the literature for randomly oriented Platonic solids~\cite{Stanley,Torquato,Torquato2,Hyytia,koza2} as well as aligned cubes.
Density thresholds in the case of grain-grain 
percolation ($\eta_{c1}$) and sheath tunnel ($\eta_{c*}$) transitions invariably occur for lower 
concentrations of randomly oriented solids than for the aligned counterparts, though by a comparatively 
greater margin in the case of sheath/tunnel transitions.  However, the trend is less  
consistent in the case of void percolation transitions ($\eta_{c2}$).   
As in prior results, $\eta_{c2}$ is significantly less for randomly 
oriented than aligned cubes, and though to a more subtle degree, this trend holds for tetrahedra, dodecahedra,
and the semi-regular truncated icosahedron. On 
the other hand, for both octahedra and icosahedra, $\eta_{c2}$ is less for the aligned solids than for 
randomly oriented counteparts.

Critical exponents for grain-grain and void percolation transitions are consistent with $\nu = 0.8764(12)$~\cite{wang} for the 
3D percolation universality class; on the other hand, the critical exponents $\nu_{*}$ are not compatible 
with the latter and appear to be non-universal at least with respect to randomly oriented and aligned cases.  Whereas
the critical exponents for aligned solids cluster about $\nu_{*} = 1.25$, $\nu_{*}$ results for randomly oriented solids 
are consistent with $\nu_{*} = 1.7$ up to Monte Carlo Statistical error.

\begin{figure}
\includegraphics[width=.5\textwidth]{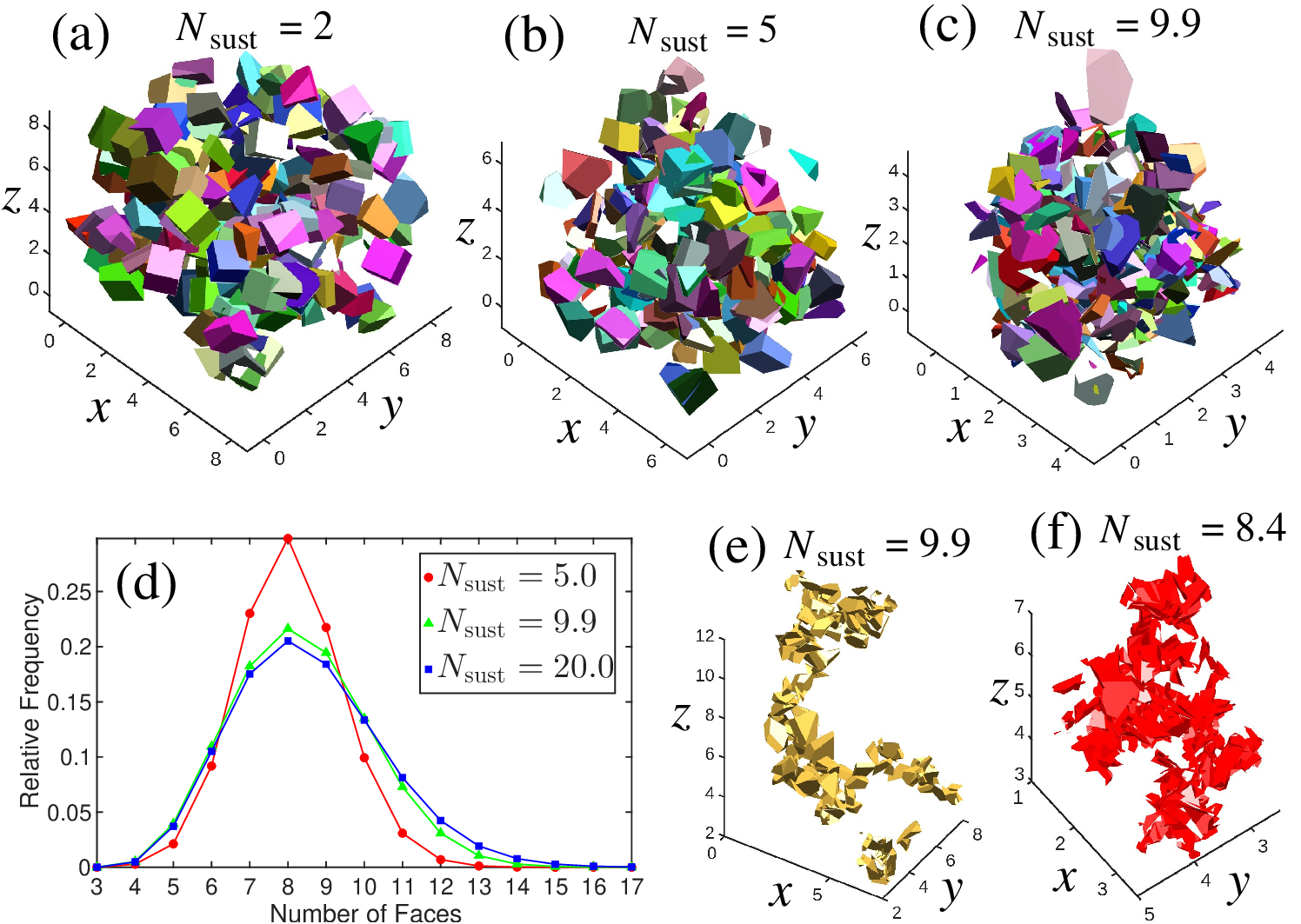}
\caption{\label{fig:Fig4} (Color online) Panel (a), (b), and (c) depict sample assemblies of 
grains for various numbers of sustained slices $N_{\mathrm{sust}}$.  The graph in panel (d) is a 
frequency plot for facet number for a range of $N_{\mathrm{sust}}$ values.  Panels (e) and (f) show
a portion of a percolating free surface near $\rho_{c1}$ and $\rho_{c2}$ for the $N_{\mathrm{sust}}$ 
values indicated.}
\end{figure}

Finally, we consider the effect of structural disorder on $\rho_{c1}$, $\rho_{c*}$, and $\rho_{c2}$;
we generate disordered fragments by
successively cleaving away material from cubes.  The number of
randomly positioned and oriented fracture planes is sampled from Poissonian statistics for a tunable 
density per unit volume.
However, as a more pertinent parameter, we characterize the 
fractured inclusions in terms of $N_{\mathrm{sust}}$, the mean number of sustained slices
(i.e. fracturing planes which cleave away at least one vertex).
The mean inclusion volume $\langle v_{\mathrm{B}} \rangle$  used, e.g. to find 
$\eta_{c} = \rho_{c} \langle v_{\mathrm{B}} \rangle$, is calculated by 
summing volumes of component tetrahedra defined by the polyhedron center, the face center, and the two vertices 
of a facet edge, and by averaging over a suitable number ($10^{9}$ in this work) of fragment realizations.

Structurally disordered fragments are generated 
efficiently by keeping track of planes, edges, and vertices as material is successively cleaved away;  
a plane is eliminated if none of its vertices remain after a sustained slice. As may be seen in panel (d) of Fig.~\ref{fig:Fig3},
after many fracture events, the distribution of the number of fragment facet tends to a limiting form peaked at eight faces.
This typically modest number of faces makes finding free surfaces computationally tractable 
even for $N_{\mathrm{sust}} \gg 1$.
In panels (a), (b), and (c) of Fig.~\ref{fig:Fig4}, assemblies of fractured inclusions are shown from fragments 
with most of the parent solid intact ($N_{\mathrm{sust}} = 2.0$) to ($N_{\mathrm{sust}} = 9.9$) 
where primordial facets are mostly eliminated.

\begin{figure}
\includegraphics[width=.45\textwidth]{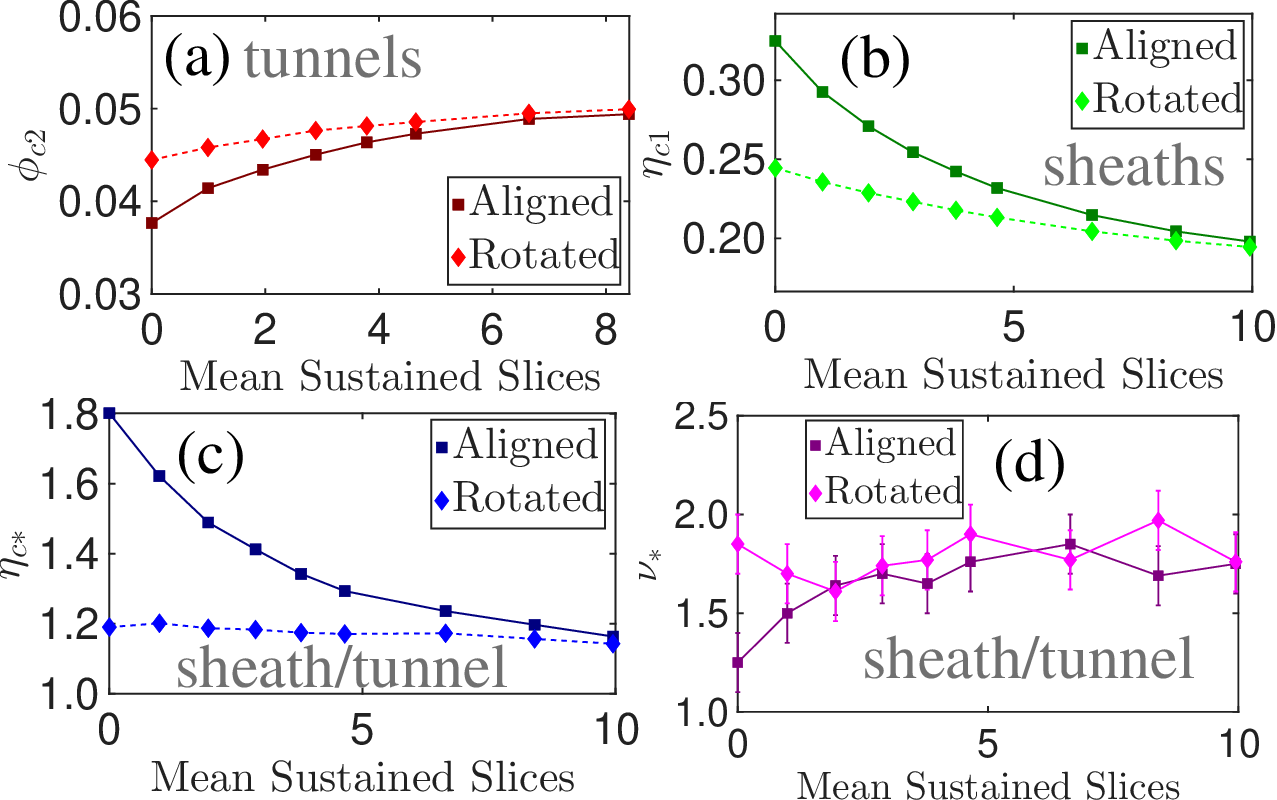} 
\caption{\label{fig:Fig5} (color online) Critical concentrations are shown in panels 
(a), (b), and (c) for $\phi_{c2} = e^{-\eta_{c2}}$ in the case of void percolation transitions, for percolation transitions 
of overlapping fragments, and for the sheath/tunnel transition respectively.  
Lines are guides to the eye, and Monte Carlo statistical errors are smaller than the symbol sizes.  
Panel (d) shows the critical
exponent $\nu_{*}$ with respect to $N_{\mathrm{sust}}$.}
\end{figure}

Figure~\ref{fig:Fig5} shows critical indices for the grain-grain percolation transition, the sheath to tunnel transition, and 
void percolation. In each case, $\eta_{c}$ is calculated both for aligned cube shaped grains and 
randomly oriented cubes that are then fractured with convergence of the critical concentrations beyond on the 
order of ten sustained slices.   In the case of $\eta_{c2}$, the critical fractional volume
(i.e. $\phi_{cc} = e^{-\eta_{2c}}$) is shown in panel (a) of Fig.~\ref{fig:Fig4}.  The critical porosity 
curves for the randomly oriented and aligned systems converge to $\phi_{c} = 0.050(1)$.
Critical concentrations for the percolation transitions 
of overlapping fragments displayed in panel (b) of Fig.~\ref{fig:Fig4} also converge to a common $\eta_{c1}$ value 
in the large $N_{\mathrm{sust}}$ limit. A similar convergence is evident in the plot showing $\eta_{c*}$ for the sheath/tunnel 
transition in panel (c) of Fig.~\ref{fig:Fig4}.  A salient feature of sheath/tunnel transition 
in the case of randomly oriented grains 
is the near-constancy of $\eta{c*}$ with respect to $N_{\mathrm{sust}}$.  Finally, panel (d) is a plot of the critical 
exponent $\nu_{*}$; with the exception of $N_{\mathrm{sust}} = 0$, the $\nu_{*}$ exponents are within each other's error bars 
for aligned grains and the randomly oriented counterparts. 

In conclusion, we have developed a technique for finding free surfaces and determining if they percolate and if they are 
sheath-like or tunnel-like.  With large-scale Monte Carlo calculations, we have significantly improved the 
accuracy of critical concentrations for void percolation in the case of the Platonic solids as well
as the semi-regular truncated icosahedron. 
We posit the existence of a third phase boundary at $\rho_{c*}$ where,
in the bulk limit, free surfaces bounding grains abruptly become exposed surfaces lining tunnels. 


\begin{acknowledgments}
We acknowledge helpful discussions with Michael Crescimanno.
\end{acknowledgments}


\end{document}